\documentclass[12pt]{article}
\usepackage{amsmath,amsfonts,amssymb}

\begin{document}
\newcommand{\todo}[1]{{\em \small {#1}}\marginpar{$\Longleftarrow$}}   
\newcommand{\labell}[1]{\label{#1}\qquad_{#1}} 

\rightline{DCPT-09/73}   
\vskip 1cm 

\begin{center} {\Large \bf Black holes and black strings in plane waves}
\end{center} 
\vskip 1cm   
  
\renewcommand{\thefootnote}{\fnsymbol{footnote}} 
\centerline{\bf Julian Le Witt\footnote{j.a.le-witt@durham.ac.uk} and
  Simon F. Ross\footnote{S.F.Ross@durham.ac.uk} }
\vskip .5cm 
\centerline{\it Centre for Particle Theory, Department of Mathematical
  Sciences} 
\centerline{\it Durham University, South Road, Durham DH1 3LE, U.K.}

\setcounter{footnote}{0}   
\renewcommand{\thefootnote}{\arabic{footnote}}

\begin{abstract}
  We investigate the construction of black holes and black strings in
  vacuum plane wave spacetimes using the method of matched asymptotic
  expansions. We find solutions of the linearised equations of motion
  in the asymptotic region for a general source on a plane wave
  background. We observe that these solutions do not satisfy our
  previously defined conditions for being asymptotically plane
  wave. Hence, the space of asymptotically plane wave solutions is
  restricted. We consider the solution in the near region, treating
  the plane wave as a perturbation of a black object, and find that
  there is a regular black string solution but no regular black hole
  solution.
\end{abstract}

\section{Introduction}

Plane waves are of interest both from the point of view of classical
gravity and in the context of considerations of holography in string
theory. To a relativist, the plane waves are a rich class of exact
solutions, which can be obtained as the result of applying the Penrose
limit to an arbitrary spacetime. In string theory, they are of
interest because the theory on the worldsheet admits a simple
realisation, making explicit computations possible
\cite{Metsaev:2001bj}. Secondly, certain maximally supersymmetric
plane wave backgrounds \cite{Blau:2001ne} admit a dual field theory
interpretation as a scaling limit of certain field theories
\cite{Berenstein:2002jq}. This correspondence is obtained by taking a
Penrose limit of the AdS/CFT correspondence. As a result, our
understanding of holography in this case is rather indirect, and no
holographic dictionary has yet been constructed for this duality.

From both these points of view, the study of black hole solutions with
plane wave asymptotics is clearly interesting. Black holes for other
``simple'' asymptotics, such as flat space or anti-de Sitter space,
have long been known. The construction of black hole solutions with
plane wave asymptotics would offer a new, rich family of black hole
solutions, whose thermodynamics could exhibit interesting dependence
on the asymptotic plane wave considered. For string theory, the black
hole solutions in the maximally supersymmetric plane wave
backgrounds of \cite{Blau:2001ne} would presumably be related to
finite-temperature excitations of the corresponding field theory, as
in the AdS/CFT correspondence \cite{Witten:1998zw}. Consideration of
such solutions could therefore cast interesting light on this
poorly-understood duality.

Some exact solutions describing black strings in plane wave
backgrounds have been obtained by applying solution generating
transformations
\cite{Kaloper:1996hr,Hubeny:2002pj,Hubeny:2002nq,Gimon:2003ms,Gimon:2003xk}. A
review of this work and the structure of horizons and plane waves can
be found in \cite{Hubeny:2003ug}. However, such methods are available
only in special cases, and a solution describing the simplest
situation, a regular black hole or black string in a vacuum plane wave
background, has not been obtained by these methods. Constructing
solutions by directly solving the equations of motion is challenging.

In this paper, we adopt the method of matched asymptotic expansions to
find approximate solutions when the horizon size $r_+$ of the black
hole or black string is small compared to the curvature scale
$\mu^{-1}$ of the plane wave. This gives a separation of scales, which
can be exploited to solve the equations of motion in the linearised
approximation in separate regions, matching the solutions in an
overlap region. Such methods have been successfully applied to the
construction of caged black holes in Kaluza-Klein theory
\cite{Harmark:2003yz,Gorbonos:2004uc} and to construct black ring
solutions in more than five spacetime dimensions \cite{Emparan:2007wm}
and in anti-de Sitter space \cite{Caldarelli:2008pz}. These ideas have
been further developed in \cite{Emparan:2009cs,Emparan:2009at}, where
general extended black objects wrapping a submanifold in an arbitrary
spacetime have been considered at leading order in the region far from
the black object.

We proceed in a similar way to these previous examples, first finding
the metric far from the source (for $ r \gg r_+$) by studying the
linearised approximation to gravity with an appropriate delta-function
source. The wave equation in the plane wave background is rather
complicated, so we focus on solving this problem in an intermediate
region $ r_+ \ll r \ll \mu^{-1}$ where the deviations from flat space
due to both the source and the plane wave are small.

Solving the equation in this regime, we find that simple dimensional
analysis indicates that the solutions will violate the asymptotic
boundary conditions proposed in \cite{LeWitt:2008zx} as a definition
of asymptotically plane wave spacetimes. In fact, the perturbation due
to the delta-function source becomes large relative to the background
metric at large distances. An explicit analysis in four and five
dimensions shows that the terms violating these boundary conditions
are indeed non-zero. These solutions thus appear not to be
asymptotically plane wave; we will refer to them as black holes or
black strings in plane wave backgrounds. The fact that the linearised
solutions for a delta-function source violate the asymptotic boundary
conditions suggests that as in AdS$_2$ \cite{Maldacena:1998uz} and the
Kerr/CFT correspondence \cite{Guica:2008mu,Amsel:2009ev,Dias:2009ex},
the space of asymptotically plane wave spacetimes may be highly
restricted.

We then obtain the near horizon metric in the region $r \ll \mu^{-1}$
by solving the linearised Einstein equations on the background of the
black object, treating the plane wave as a perturbation. For a black
hole, we find that there is no linearised solution which is regular on
the horizon. For the black string, we obtain a regular solution in the
near region, and verify that it matches on to the solution in the
intermediate region.

When solving the equations, we focus on vacuum plane waves in the
lowest possible dimension for simplicity, but the method of matched
asymptotic expansion is more general, and a similar analysis could be
applied to construct black string solutions in any plane wave
background of interest in arbitrary dimensions. We will remark on the
extension to other waves and higher dimensions at appropriate points
in the calculation. The calculation in the region $r \gg r_+$ is
described in section \ref{int}, and the calculation in the region $r
\ll \mu^{-1}$ is described in section \ref{near}. We conclude with
some remarks on the implications of our results in section
\ref{concl}. 

\section{Linearised solutions on a plane wave background}
\label{int}

We want to construct solutions corresponding to a black hole or black
string of radius $r_+$ in a general vacuum plane wave background in $D=d+2$
dimensions
\begin{equation}
ds^{2}=-dt^{2}+dz^{2}+\mu_{\alpha\beta}(t+z)x^{\alpha}x^{\beta}(dt+dz)^{2}+\delta_{\alpha\beta}dx^{\alpha}dx^{\beta},\label{eq:GPW_Ddim}  
\end{equation}
where $x^\alpha$, $\alpha = 1, \ldots d$ are Cartesian coordinates on
the transverse space. We will work in the parameter range $r_+ \ll
\mu^{-1}$, where we take the matrix $\mu_{\alpha\beta}(t+z)$
characterising the wave to have a single characteristic scale $\mu$
for simplicity. The black object can then be treated as a small
perturbation of the plane wave background for $r \gg r_+$. In this
region of the spacetime, the problem of constructing a black hole or
black string solution thus reduces to solving the linearised
Einstein's equations for a suitable source $T_{\mu\nu}$. In transverse
gauge, the linearised equations are\footnote{Note that in our actual
  calculations, we will not assume the transverse traceless gauge, as
  it is more convenient to use the gauge freedom to fix particular
  components of the perturbation.}
\begin{equation} \label{lineq}
\Box \bar{h}_{\mu\nu} = -16 \pi G T_{\mu\nu}.
\end{equation}
For a pointlike
source, the relevant stress tensor is simply $T_{\mu\nu} = M V_\mu
V_\nu \delta(x^\mu - x^\mu(\tau))$, where $x^\mu(\tau)$ is the
particle's trajectory, $V^\mu = dx^\mu/d\tau$ is the tangent to this
trajectory, and $M$ is the proper mass. For a black string solution,
the stress tensor can be determined by linearising the vacuum black
string solution in $d+2$ dimensions,
\begin{equation} \label{bsmet}
ds^2 = -\left( 1 - \frac{r_+^{d-2}}{r^{d-2}} \right) dt^2 + dz^2 + \left( 1 -
  \frac{r_+^{d-2}}{r^{d-2}} \right)^{-1} dr^2 + r^2 d\Omega_{d-1}^2,
\end{equation}
which gives the stress tensor in these coordinates as
\begin{equation} \label{bsstress}
T_{tt} = \frac{(d-1) r_+^{d-2}}{16 \pi G}\delta^{d}(r), \quad T_{zz} =
- \frac{r_+^{d-2}}{16 \pi G}\delta^{d}(r).
\end{equation}

The source is fixed to follow some appropriate trajectory in the plane
wave background. For a pointlike source, the appropriate trajectory is
a timelike geodesic of the background spacetime. To obtain a
stationary black hole solution, we should require this geodesic to be
the orbit of a timelike Killing vector in the spacetime. This forces
us to restrict to plane waves with a constant matrix
$\mu_{\alpha\beta}(t+z) = \mu_{\alpha\beta}$, so that the solution has
a timelike Killing vector, and to consider the geodesic
$z=0$,\footnote{We can make this choice without loss of generality by
  translation invariance in $z$.} $x^\alpha = 0$, which is the unique geodesic
trajectory which is also an orbit of the Killing vector. The
appropriate source is then $T_{tt} = M \delta(z) \delta^d(x^\alpha)$,
and the size of the black hole is  $r_+^{d-1} \propto M$.

For the black string, the equation of motion for a probe string
is \cite{Carter:2000wv}
\begin{equation}
K_{\mu\nu}^{\ \ \rho} T^{\mu\nu} = 0,
\end{equation}
where $K_{\mu\nu}^{\ \ \rho}$ is the second fundamental tensor of the
submanifold defining the embedding of the string worldvolume, and
$T_{\mu\nu}$ is the stress tensor of the source. We will consider
embedding the black string along the submanifold $x^\alpha =0$, which
has $K_{\mu\nu}^{\ \ \rho} = 0$. As a result, there is no constraint
on the form of the stress tensor. As for the black hole, we need to
restrict to constant $\mu_{\alpha\beta}(t+z) = \mu_{\alpha\beta}$ so
that this submanifold is an orbit of the spacetime isometries, so that
we can expect to obtain a stationary uniform black string solution. We
can then use boosts in the $t-z$ plane to choose the black string
solution to be in its rest frame, setting $T_{tz}=0$, without loss of
generality. The appropriate source is thus \eqref{bsstress}. We want
to find a uniform black string solution, so the components of the
stress tensor are assumed to be constants along the worldvolume. The
blackfold equations of \cite{Emparan:2009at} are hence trivially
satisfied.

In each case, the problem thus reduces in principle to solving
\eqref{lineq} on the plane wave background for an appropriate
source. However, we do not have the Green's function for this
differential equation in closed form, so we will content ourselves
with studying this problem in the intermediate region $r_+ \ll r \ll
\mu^{-1}$, where we can treat the plane wave itself as a small
perturbation of flat space, and obtain the solution of \eqref{lineq}
order by order in $\mu^2 r^2$.

\subsection{Dimensional analysis}

We first discuss the perturbation in general dimensions using a simple
dimensional analysis argument. For the case of a point source, we find
it convenient to rewrite the metric in spherical polar coordinates,
introducing a radial coordinate
\begin{equation}
r^2 = z^2 + \delta_{\alpha\beta} x^{\alpha} x^{\beta}, 
\end{equation}
and defining coordinates $\theta^i$ on the $S^d$ at constant $r$. As
in \cite{Kodama:2003jz}, we use $a,b$ to denote coordinates on the two
dimensional space spanned by $r,t$. By dimensional analysis, the form
of the perturbation to first order in $M$ and in $\mu^2$ will be
\begin{eqnarray}
h_{ab} & = &
\frac{M}{r^{D-3}}h_{ab}^{(0)}+\frac{M\mu^{2}}{r^{D-5}}h_{ab}^{(1)}(\theta^i),\nonumber
\\
h_{ai} & = &
\frac{M}{r^{D-4}}h_{ai}^{(0)}+\frac{M\mu^{2}}{r^{D-6}}h_{ai}^{(1)}(\theta^i),\label{eq:h_BH_int_Ddim}
\\
h_{ij} & = &
\frac{M}{r^{D-5}}h_{ij}^{(0)}+\frac{M\mu^{2}}{r^{D-7}}h_{ij}^{(1)}(\theta^i),
\nonumber
\end{eqnarray}
where $h^{(0)}_{\mu\nu}$ and $h^{(1)}_{\mu\nu}$ are dimensionless
functions depending only on the angles $\theta^i$. In fact, since the
spherical symmetry is only broken by the plane wave, $h^{(0)}_{ab}$
are constants, and the component on the sphere $h^{(0)}_{ij}$ will be
proportional to the metric on the sphere $\gamma_{ij}$ We will always
work in a gauge where $h^{(0)}_{ij}$ vanishes. Each addition of an $i$
index raises the power of $r$ by one because the coordinates on the
sphere are written in terms of dimensionless angles.

This simple dimensional analysis already indicates a significant
issue: this perturbation does not satisfy the boundary conditions
introduced in \cite{LeWitt:2008zx}. There, it was assumed that
components of the perturbation in the directions transverse to the
wave would fall off at least as $1/r^{D-4}$ (corresponding to $h_{ij}
\propto 1/r^{D-6}$, because of the extra factors of $r$ from writing
the perturbation in polar coordinates), characteristic of a localised
source in a flat spacetime. However, we find that the term resulting
from the interaction with the wave must grow more quickly than this on
dimensional grounds. When we think of the plane wave as a perturbation
around flat space, the plane wave background introduces corrections
which grow more quickly with $r$ than the original leading-order
response.

Similarly, when we consider a black string source, it is convenient to
write the metric in the directions transverse to the wave in polar
coordinates, introducing a radial coordinate 
\begin{equation}
r^2 = \delta_{\alpha\beta} x^{\alpha} x^{\beta}, 
\end{equation}
and introducing coordinates $\theta^{i}$ on the $S^{d-1}$ at constant
$r,z$. In the string source case, $a,b$ will denote coordinates in the
three dimensional space spanned by $t,r,z$. Then to leading order in
$r_+$ and $\mu^2$, the perturbation sourced by a black string will
have the form
\begin{align}
h_{ab} & =\frac{r_+^{D-4}}{r^{D-4}}h_{ab}^{(0)}+\frac{r_+^{D-4}
  \mu^{2}}{r^{D-6}}h_{ab}^{(1)}(\theta^i), \nonumber \\
h_{ai} & =\frac{r_+^{D-4}}{r^{D-5}}h_{ai}^{(0)}+\frac{r_+^{D-4}\mu^{2}}{r^{D-7}}h_{ai}^{(1)}(\theta^i),\\
h_{ij} &
=\frac{r_+^{D-4}}{r^{D-6}}h_{ij}^{(0)}+\frac{r_+^{D-4}\mu^{2}}{r^{D-8}}h_{ij}^{(1)}(\theta^i), \nonumber
\end{align}
where $h^{(0)}_{ab}$ are constants and $h^{(1)}_{\mu\nu}$ are
functions of the coordinates $\theta^i$ on the sphere only. Thus, as
in the black hole case, the perturbation does not satisfy the boundary
conditions introduced in \cite{LeWitt:2008zx}.

This is a significant issue because at least in low spacetime
dimensions, the resulting perturbation actually grows more quickly
with $r$ than the background metric. In $D=4$ for the black hole and
$D=5$ for the black string, the perturbation of the angular metric
$h_{ij}$ has a contribution that goes like $r_+ \mu^2 r^3$, which is
growing faster than the background metric on the sphere, which goes
like $r^2$. Furthermore, what we have discussed so far is just the
leading order correction in $\mu^2$. Higher order terms in $\mu^2$
will come with additional powers of $r$. One might hope that when the
problem is solved to all orders in $\mu^2$, the resulting behaviour
could be under better control, but it is hard to see how such a
cancellation between different orders could be arranged. We will see
later in a particular example that this does not occur.

Thus, we are faced with the odd situation that the linearised field of
a point source may become more important than the background,
signalling a breakdown of the linearised approximation, far from the
source itself. Thus, the solutions we construct should not be thought
of as ``asymptotically plane wave'' black holes/strings, as the metric
in the asymptotic regime is not close to the original plane wave
metric. As a result, the analysis of \cite{LeWitt:2008zx} will not
apply to these spacetimes, and in particular we do not expect that
they will have finite action with respect to the action principle
discussed there.

One might hope that the terms which violate those boundary conditions
which are allowed by dimensional analysis may actually vanish. This
hope would be encouraged by the fact that the specific examples of
plane wave black strings constructed in
\cite{Kaloper:1996hr,Hubeny:2002pj,Hubeny:2002nq} satisfied the
asymptotic boundary conditions of \cite{LeWitt:2008zx}.  However, the
examples of \cite{Kaloper:1996hr,Hubeny:2002pj,Hubeny:2002nq} are
special cases in that they are constructed by the Garfinkle-Vachaspati
solution-generating transformation \cite{Garfinkle:1990jq}, and by
construction can only differ from the seed solution in the metric
components along the null direction. By contrast, the solution
constructed in \cite{Gimon:2003xk}, which was obtained by a different
method, has precisely the kinds of corrections that are predicted by
this dimensional analysis argument.

In the next two subsections, we will consider the solution of the
linearised equations of motion for the perturbation in detail for the
lowest possible dimension for black hole and black string sources, and
see in these particular examples that the terms which violate the
asymptotic boundary conditions of \cite{LeWitt:2008zx} do indeed
appear. Thus, the approximate solutions we obtain for black holes and
black strings in plane wave backgrounds are not asymptotically plane
wave in the sense defined in \cite{LeWitt:2008zx}. Given the above
dimensional analysis arguments and the results below, it seems
reasonable to expect that this is the generic case, so that the space
of asymptotically plane wave solutions is very limited. We will
comment on this in the conclusions.

\subsection{Black hole}

Let us consider the perturbation sourced by a point source in the
lowest possible dimension, $D=4$, in detail. By a choice of
coordinates, the most general four dimensional vacuum plane wave can be
written as
\begin{equation} \label{4dpw}
ds_{wave}^{2}=-dt^{2}+dx^{2}+dy^{2}+dz^{2}-\mu^{2}(x^{2}-y^{2})(dt+dz)^{2}.
\end{equation}
We rewrite this in spherical polars by defining 
\begin{equation}
z = r \cos \theta, \quad x = r \sin \theta \cos \phi, \quad y = r \sin
\theta \sin \phi,  
\end{equation}
so
\begin{eqnarray}
ds_{wave}^{2}&=&-dt^{2}+dr^2+r^2(d\theta^2 + \sin^2 \theta d\phi^2) \\
&&-\mu^{2}r^2 \sin^2 \theta (\cos^2\phi - \sin^2 \phi) (dt+\cos \theta
dr - r \sin \theta d\theta)^{2}. \nonumber
\end{eqnarray}

As in the above subsection, we can use dimensional analysis to fix the
dependence of the perturbation on $r$. We can in fact determine the
perturbation to zeroth order in $\mu^2$ by simply linearising the
Schwarzschild solution, which gives $h_{tt} = h_{rr} =
\frac{2M}{r}$. This satisfies the linearised equations of motion for a
delta-function point source, but not in the transverse traceless gauge
which was assumed in writing \eqref{lineq}. In what follows, we will
not assume the transverse traceless gauge, as it is more convenient to
use the gauge freedom to fix some components of the perturbation.

For the terms of first order in $\mu^2$, we can use the freedom to
choose a gauge for the perturbation to set $h^{(1)}_{a \phi}$ and
$h^{(1)}_{\theta \phi}$ to zero. Note that we have four gauge degrees
of freedom but have only eliminated three components, hence we have
one remaining degree of freedom which we will use later.  We then make
an ansatz for the $\phi$ dependence of the perturbation, and write our
perturbation as
\begin{eqnarray}
h_{ab} & = &
\frac{M}{r}h_{ab}^{(0)}+M\mu^{2}r(\cos^{2}\phi-\sin^{2}\phi)h_{ab}^{(1)}(\theta),
\nonumber \\
h_{a\theta} & = & M\mu^{2}r^{2}(\cos^{2}\phi-\sin^{2}\phi)h_{a\theta}^{(1)}(\theta),\\
h_{ij} & = &
M\mu^{2}r^{3}(\cos^{2}\phi-\sin^{2}\phi)h_{ij}^{(1)}(\theta), \nonumber
\end{eqnarray}
where the non-zero components of $h_{ab}^{(0)}$ are
$h_{tt}^{(0)}=2,h_{rr}^{(0)}=2$, and the non-zero components of
$h_{\mu\nu}^{(1)}(\theta)$ are $h_{tt}^{(1)}(\theta)$, $h_{tr}^{(1)}(\theta)$,
$h_{t\theta}^{(1)}(\theta)$, $h_{rr}^{(1)}(\theta)$,
$h_{r\theta}^{(1)}(\theta)$, $h_{\theta\theta}^{(1)}(\theta)$ and
$h_{\phi\phi}^{(1)}(\theta)$. 

We now want to substitute this ansatz into the linearised Einstein
equations and solve for the undetermined functions
$h_{\mu\nu}^{(1)}(\theta)$, requiring regularity on the sphere. In an
arbitrary gauge, the linearised Einstein equations for $r \neq 0$ are
\begin{equation}
R_{\mu\nu}^{(1)}=\frac{1}{2}g^{\rho\sigma}(\nabla_{\rho}\nabla_{\mu}h_{\nu\sigma}+\nabla_{\rho}\nabla_{\nu}h_{\mu\sigma}-\nabla_{\mu}\nabla_{\nu}h_{\rho\sigma}-\nabla_{\rho}\nabla_{\sigma}h_{\mu\nu})=0.\label{eq:EinsteinEqLinear}
\end{equation}
Substituting our ansatz, these equations become (where primes denote
derivatives with respect to $\theta$)
\begin{eqnarray}
-\sin^{2}\theta h_{tt}^{(1)\prime\prime}(\theta)-\sin\theta\cos\theta h_{tt}^{(1)\prime}(\theta)+2(\cos^{2}\theta+1)h_{tt}^{(1)}(\theta)\label{eq:BHFR_tt}\\
-6\cos^{6}\theta-2\cos^{4}\theta+22\cos^{2}\theta-14=0,\nonumber 
\end{eqnarray}
\begin{eqnarray}
-\sin^{2}\theta h_{tr}^{(1)\prime\prime}(\theta)-\sin\theta\cos\theta h_{tr}^{(1)\prime}(\theta)+2\sin^{2}\theta h_{t\theta}^{(1)\prime}(\theta)+4h_{t\theta}^{(1)}(\theta)\label{eq:BHFR_tr}\\
+2\sin\theta\cos\theta
h_{t\theta}^{(1)}(\theta)-8\cos^{5}\theta+16\cos^{3}\theta-8\cos\theta=0,\nonumber 
\end{eqnarray}
\begin{eqnarray}
-\sin^{2}\theta h_{rr}^{(1)\prime\prime}(\theta)-\sin\theta\cos\theta h_{rr}^{(1)\prime}(\theta)+4\sin^{2}\theta h_{r\theta}^{(1)\prime}(\theta)+2(3-\cos^{2}\theta)h_{rr}^{(1)}(\theta)\label{eq:BHFR_rr}\\
-2\sin^{2}\theta h_{\theta\theta}^{(1)}(\theta)-2\sin^{2}\theta
h_{\phi\phi}^{(1)}(\theta)+10\cos^{6}\theta-26\cos^{4}\theta+22\cos^{2}\theta-6=0,\nonumber 
\end{eqnarray}
\begin{eqnarray}
\sin^{2}\theta h_{rr}^{(1)\prime}(\theta)-\sin^{2}\theta h_{\phi\phi}^{(1)\prime}(\theta)-\sin\theta\cos\theta h_{\phi\phi}^{(1)}(\theta)+\sin\theta\cos\theta h_{\theta\theta}^{(1)}(\theta)\label{eq:neq5}\\
+2(\cos^{2}\theta+1)h_{r\theta}^{(1)}(\theta)-4\sin\theta(\cos^{5}\theta-2\cos^{3}\theta+\cos\theta)=0,\nonumber 
\end{eqnarray}
\begin{equation}
\sin^{2}\theta
h_{tr}^{(1)\prime}(\theta)+2(\cos^{2}\theta+1)h_{t\theta}^{(1)}(\theta)+2\sin\theta(\cos^{4}\theta-4\cos^{2}\theta+3)=0,\label{eq:neq4}
\end{equation}
\begin{equation}
h_{r\theta}^{(1)\prime}(\theta)-\cot\theta
h_{r\theta}^{(1)}(\theta)-h_{\theta\theta}^{(1)}(\theta)+h_{rr}^{(1)}(\theta)-\cos^{4}\theta+2\cos^{2}\theta-1=0,\label{eq:neq3}
\end{equation}
\begin{equation}
h_{t\theta}^{(1)\prime}(\theta)-\cot\theta
h_{t\theta}^{(1)}(\theta)+h_{tr}^{(1)}(\theta)+2\cos\theta(1-\cos^{2}\theta)=0,\label{eq:neq2}
\end{equation}
\begin{equation}
-\sin\theta(h_{rr}^{(1)\prime}(\theta)-h_{tt}^{(1)\prime}(\theta))+\cos\theta(h_{rr}^{(1)}(\theta)-h_{tt}^{(1)}(\theta))+2\sin\theta
h_{r\theta}^{(1)}(\theta)+2\cos\theta(\cos^{4}\theta-1)=0,\label{eq:neq1}
\end{equation}
\begin{eqnarray}
-\sin^{2}\theta h_{\phi\phi}^{(1)\prime\prime}(\theta)+\sin^{2}\theta h_{tt}^{(1)\prime\prime}(\theta)+\sin\theta\cos\theta(h_{\theta\theta}^{(1)\prime}(\theta)-2h_{\phi\phi}^{(1)\prime}(\theta))\nonumber \\
-\sin^{2}\theta h_{rr}^{(1)\prime\prime}(\theta)6\sin^{2}\theta h_{r\theta}^{(1)\prime}(\theta)+-5\sin^{2}\theta h_{\theta\theta}^{(1)}(\theta)+3\sin^{2}\theta h_{rr}^{(1)}(\theta)\label{eq:BHFR_thth}\\
-\sin^{2}\theta h_{\phi\phi}^{(1)}(\theta)+\sin^{2}\theta
h_{tt}^{(1)}(\theta)+2\cos^{2}\theta(\cos^{4}\theta+3\cos^{2}\theta-5)+2=0,\nonumber 
\end{eqnarray}
\begin{eqnarray}
\sin^{2}\theta h_{\phi\phi}^{(1)\prime\prime}(\theta)+\sin\theta\cos\theta(h_{rr}^{(1)\prime}(\theta)-h_{tt}^{(1)\prime}(\theta)+2h_{\phi\phi}^{(1)\prime}(\theta)-h_{\theta\theta}^{(1)\prime}(\theta))\nonumber \\
+\cos^{2}\theta(3h_{rr}^{(1)}(\theta)-3h_{\theta\theta}^{(1)}(\theta)+h_{tt}^{(1)}(\theta))-2\sin^{2}\theta h_{r\theta}^{(1)\prime}(\theta)+3\sin^{2}\theta h_{\phi\phi}^{(1)}(\theta)\label{eq:BHFR_pp}\\
-7h_{rr}^{(1)}(\theta)+3h_{tt}^{(1)}(\theta)+2\cos^{2}\theta(3\cos^{4}\theta-7\cos^{2}\theta+9)-10=0.\nonumber 
\end{eqnarray}
We have a system of ten equations in seven unknown functions (in fact,
there will be only six unknown functions once we have made use of
the one remaining degree of gauge freedom) and so it seems our system
is over-constrained. We find, however, that there are only six independant
equations and hence that our system is in fact well defined. It is
convenient to subtract a multiple of \eqref{eq:neq3} from \eqref{eq:BHFR_rr}
to simplify it to
\begin{eqnarray}
-\sin^{2}\theta h_{rr}^{(1)\prime\prime}(\theta)-\sin\theta\cos\theta h_{rr}^{(1)\prime}(\theta)+2(1+\cos^{2}\theta)h_{rr}^{(1)}(\theta)+8\sin\theta\cos\theta h_{r\theta}^{(1)}(\theta)\label{eq:neq6}\\
+2\sin^{2}\theta h_{\theta\theta}^{(1)}(\theta)-2\sin^{2}\theta
h_{\phi\phi}^{(1)}(\theta)+10\cos^{2}\theta-14\cos^{4}\theta+6\cos^{6}\theta-2=0.\nonumber 
\end{eqnarray}
By using combinations of \eqref{eq:neq3}, \eqref{eq:neq1}, \eqref{eq:neq6}
and their derivatives it is possible to reduce \eqref{eq:BHFR_tt}
to an algebraic equation
\begin{eqnarray}
2\sin\theta\cos\theta h_{r\theta}^{(1)}(\theta)+2\sin^{2}\theta h_{\phi\phi}^{(1)}(\theta)+3h_{tt}^{(1)}(\theta)-5h_{rr}^{(1)}(\theta)\label{eq:neq5a}\\
+2\cos^{2}\theta
h_{rr}^{(1)}(\theta)+2(-\cos^{6}\theta+4\cos^{4}\theta+\cos^{2}\theta-4)=0.\nonumber 
\end{eqnarray}

We find we can write \eqref{eq:BHFR_tr}, \eqref{eq:neq5},
\eqref{eq:BHFR_thth} and \eqref{eq:BHFR_pp} as linear combinations of
\eqref{eq:neq4}, \eqref{eq:neq3}, \eqref{eq:neq2}, \eqref{eq:neq1},
\eqref{eq:neq6} and \eqref{eq:neq5a} and hence that these equations
are not independant.  We now see that a convenient choice of gauge is
one in which $h_{r\theta}^{(1)}(\theta)=0.$ In this gauge, the
solution which is regular on the sphere is 
\begin{eqnarray}
h_{\mu\nu} dx^\mu dx^\nu & = &
\frac{2M}{r}dt^{2}+\frac{2M}{r}dr^{2}+M\mu^{2}r \sin^2 \theta
(\cos^{2}\phi-\sin^{2}\phi)\times\nonumber \\ 
 &  &
 \left[(4-\frac{1}{3}\sin^{2}\theta)dt^{2}+4 \cos\theta
   dtdr-4r\sin\theta
   dtd\theta  \right. \\ \nonumber && \left. +\frac{1}{3} \sin^2
   \theta dr^2 -\frac{2}{3}r^{2}\sin^{2}\theta(d\theta^{2}+\sin^{2}\theta
   d\phi^{2})\right].
\end{eqnarray}

We note that as stated earlier, the regular solution for the terms of
first order in $\mu^2$ has non-zero components on the sphere which
grow faster than the background metric on the sphere. These solutions
are hence not asymptotically plane wave. While this leading order term
would not grow faster than the background metric in higher dimensions,
higher order terms in $\mu^2$ will in principle do so.

\subsection{Black string}
\label{intbs}

We now consider the perturbation for a black string source in the
lowest possible dimension, which is $D=5$ for the black string. The
most general vacuum plane wave solution in five dimensions is
\begin{equation} \label{5dwave}
ds_{wave}^{2}=-dt^{2}+dx^{2}+dy^{2}+dz^{2}+dw^{2}-\mu^{2}(\alpha(x^{2}+y^{2}-2w^{2})+\beta(x^{2}-y^{2}))(dt+dz)^{2};
\end{equation}
note that there is a two-parameter family of plane wave solutions
here. We rewrite this in spherical polars in the directions transverse
to the wave by writing
\begin{equation}
x = r \sin \theta \cos \phi, \quad y = r \sin \theta \sin \phi, \quad
w = r \cos \theta, 
\end{equation}
so
\begin{eqnarray}
ds_{wave}^{2}&=&-dt^{2}+dz^{2}+dr^2+r^2(d\theta^2+\sin^2 \theta d\phi^2)
\\ && -\mu^{2}(\alpha r^2(1-3 \cos^2 \theta)+\beta r^2 \sin^2 \theta (\cos^2
\phi - \sin^2 \phi))(dt+dz)^{2}. \nonumber
\end{eqnarray}

As in the previous subsection, we can determine the perturbation to
zeroth order in $\mu^2$ by simply linearising the Schwarzschild black
string solution \eqref{bsmet}, which gives $h_{tt} = h_{rr} =
\frac{2M}{r}$. We will again find it convenient to fix the gauge by
choosing some components of the perturbation to vanish at each order
in $\mu^2$. For the terms of first order in $\mu^2$, we note that the
background has an invariance under $t \to -t$, $z \to -z$ which is not
broken by the source, so the $h_{t\mu}$, $h_{z\mu}$ components for
$\mu \neq t,z$ will automatically vanish. 

At first order in $\mu^2$, we can treat the two different components
of the plane wave separately. We first consider the first-order terms
in the perturbation associated to $\alpha$. Let us therefore set
$\alpha=1$ and $\beta=0$ in the plane wave background
\eqref{5dwave}. There is then a translation invariance in $\phi$ and a
symmetry under $\phi \to -\phi$, which imply that $h_{\phi\mu}$ vanish
for $\mu \neq \phi$. We will make a choice of gauge to set
$h^{(1)}_{rr}$ and $h^{(1)}_{r\theta}$ to zero. This gauge choice
proves to be convenient for comparing to the solution in the near
region to be obtained later. The form of the perturbation is then
\begin{eqnarray}
h_{ab} & = & \frac{M}{r}h_{ab}^{(0)}+M\mu^{2}r h_{ab}^{(1)}(\theta),
\nonumber \\
h_{a\theta} & = & M\mu^{2}r^{2} h_{a\theta}^{(1)}(\theta),\\
h_{ij} & = &
M\mu^{2}r^{3} h_{ij}^{(1)}(\theta), \nonumber
\end{eqnarray}
where the non-zero components of $h_{ab}^{(0)}$ are
$h_{tt}^{(0)}=2,h_{rr}^{(0)}=2$, and the non-zero components of
$h_{\mu\nu}^{(1)}(\theta)$ are $h_{tt}^{(1)}(\theta)$,
$h_{tz}^{(1)}(\theta)$, $h_{zz}^{(1)}(\theta)$, $h_{\theta\theta}^{(1)}(\theta)$ and
$h_{\phi\phi}^{(1)}(\theta)$. 

We now want to substitute this ansatz into the linearised Einstein
equations and solve for the undetermined functions
$h_{\mu\nu}^{(1)}(\theta)$, requiring regularity on the sphere. In an
arbitrary gauge, the linearised Einstein equations for $r \neq 0$ are
\begin{equation}
R_{\mu\nu}^{(1)}=\frac{1}{2}g^{\rho\sigma}(\nabla_{\rho}\nabla_{\mu}h_{\nu\sigma}+\nabla_{\rho}\nabla_{\nu}h_{\mu\sigma}-\nabla_{\mu}\nabla_{\nu}h_{\rho\sigma}-\nabla_{\rho}\nabla_{\sigma}h_{\mu\nu})=0.
\end{equation}
Substituting our ansatz, these equations become
\begin{equation}
\partial_{\theta}^{2}h_{tt}^{(1)}(\theta)+\cot\theta\partial_{\theta}h_{tt}^{(1)}(\theta)+2h_{tt}^{(1)}(\theta)+16(1-3\cos^{2}\theta)=0,\label{eq:tt}
\end{equation}
\begin{equation}
\partial_{\theta}^{2}h_{tz}^{(1)}(\theta)+\cot\theta\partial_{\theta}h_{tz}^{(1)}(\theta)+2h_{tz}^{(1)}(\theta)+12(1-3\cos^{2}\theta)=0,\label{eq:zt}
\end{equation}
\begin{equation}
\partial_{\theta}^{2}h_{zz}^{(1)}(\theta)+\cot\theta\partial_{\theta}h_{zz}^{(1)}(\theta)+2h_{zz}^{(1)}(\theta)+8(1-3\cos^{2}\theta)=0,\label{eq:zz}
\end{equation}
\begin{equation}
h_{\theta\theta}^{(1)}(\theta)+h_{\phi\phi}^{(1)}(\theta)+2(1-3\cos^{2}\theta)=0,\label{eq:rr}
\end{equation}
\begin{equation}
\tan\theta\partial_{\theta}h_{\phi\phi}^{(1)}(\theta)-h_{\theta\theta}^{(1)}(\theta)+h_{\phi\phi}^{(1)}(\theta)+6\sin^{2}\theta=0,\label{eq:theta_r}
\end{equation}
\begin{eqnarray}
\partial_{\theta}^{2}h_{tt}^{(1)}(\theta)-\partial_{\theta}^{2}h_{\phi\phi}^{(1)}(\theta)-\partial_{\theta}^{2}h_{zz}^{(1)}(\theta)+\cot\theta(\partial_{\theta}h_{\theta\theta}^{(1)}(\theta)-2\partial_{\theta}h_{\phi\phi}^{(1)}(\theta)))\label{eq:th_th}\\
+h_{tt}^{(1)}(\theta)-h_{zz}^{(1)}(\theta)-5h_{\theta\theta}^{(1)}(\theta)-h_{\phi\phi}^{(1)}(\theta)+12\sin^{2}\theta-2(1-3\cos^{2}\theta)=0,\nonumber 
\end{eqnarray}
\begin{eqnarray}
\partial_{\theta}^{2}h_{\phi\phi}^{(1)}(\theta)+\cot\theta(\partial_{\theta}h_{zz}^{(1)}(\theta)+\partial_{\theta}h_{\phi\phi}^{(1)}(\theta)-\partial_{\theta}h_{\theta\theta}^{(1)}(\theta)-\partial_{\theta}h_{tt}^{(1)}(\theta))\label{eq:phi_phi}\\
+3h_{\theta\theta}^{(1)}(\theta)+3h_{\phi\phi}^{(1)}(\theta)-h_{tt}^{(1)}(\theta)+h_{zz}^{(1)}(\theta)+2(1-3\cos^{2}\theta)=0.\nonumber 
\end{eqnarray}
We first solve equations \eqref{eq:tt},\eqref{eq:zt} and \eqref{eq:zz}
for $h_{tt}^{(1)}(\theta),h_{zt}^{(1)}(\theta)$ and $h_{zz}^{(1)}(\theta)$
respectively. We then solve for $h_{\theta\theta}^{(1)}(\theta)$
and $h_{\phi\phi}^{(1)}(\theta)$ using equations \eqref{eq:BHFR_rr}
and \eqref{eq:theta_r}. It is easy to verify that these solutions
satisfy \eqref{eq:th_th} and \eqref{eq:phi_phi}. Keeping only the
regular part of the solution, we find
\begin{equation}
h_{tt}^{(1)}(\theta)=4(1-3\cos^{2}\theta),\quad
h_{tz}^{(1)}(\theta)=3(1-3\cos^{2}\theta),\quad
h_{zz}^{(1)}(\theta)=2(1-3\cos^{2}\theta),
\end{equation}
\begin{equation}
h_{\theta\theta}^{(1)}(\theta)=-(1-3\cos^{2}\theta),\quad
h_{\phi\phi}^{(1)}(\theta)=-\sin^{2}\theta(1-3\cos^{2}\theta).
\end{equation}
As in the black hole case, we see that terms that grow faster than the
background metric at large $r$ do indeed occur.

It turns out that for this background, the linearised equations of
motion can be solved exactly by including one further term at next
order in $\mu^2$. If we take 
\begin{eqnarray}
h_{ab} & = & \frac{M}{r}h_{ab}^{(0)}+M\mu^{2}r h_{ab}^{(1)}(\theta) +
M \mu^4 r^3 h_{ab}^{(2)} (\theta), \nonumber\\
h_{a\theta} & = & M\mu^{2}r^{2} h_{a\theta}^{(1)}(\theta),\\
h_{ij} & = &
M\mu^{2}r^{3} h_{ij}^{(1)}(\theta), \nonumber
\end{eqnarray}
with $h_{\mu\nu}^{(0)}$ and $h_{\mu\nu}^{(1)}$ as given above, and
\begin{equation}
h_{tt}^{(2)} = h_{tz}^{(2)} = h_{zz}^{(2)} = \frac{1}{2} (3-30 \cos^2
\theta + 27 \cos^4 \theta),
\end{equation}
this will solve the equations to linear order in $M$ but to all orders
in $\mu^2$. This gives an approximation valid in the full far region
$r \gg M$, demonstrating that the bad asymptotic behaviour of this
solution is not resolved at higher order in $\mu^2$. 

We now consider briefly the similar analysis for the other independent
component, setting $\alpha=0$ and $\beta=1$ in the plane wave
background \eqref{5dwave}. The $\phi$ dependence in this background
restricts our ability to simplify the form of the solution by general
arguments, but the results from the previous case suggest we take an
ansatz of the form
\begin{eqnarray}
h_{ab} & = & \frac{M}{r}h_{ab}^{(0)}+M\mu^{2}r \sin^2 \theta (\cos^2 \phi - \sin^2
\phi) h_{ab}^{(1)}, \nonumber \\
h_{a\theta} & = & M\mu^{2}r^{2} \sin^2 \theta (\cos^2 \phi - \sin^2
\phi) h_{a\theta}^{(1)}, \nonumber \\
h_{\theta\theta} & = & 
M\mu^{2}r^{3} \sin^2 \theta (\cos^2 \phi - \sin^2
\phi) h_{\theta\theta}^{(1)}, \\
h_{\phi\phi} & = & 
M\mu^{2}r^{3} \sin^4 \theta (\cos^2 \phi - \sin^2
\phi) h_{\phi\phi}^{(1)}, \nonumber
\end{eqnarray}
assuming the angular dependence at first order in $\mu^2$ will
reproduce the angular dependence of the background plane wave.  The
non-zero components of $h_{ab}^{(0)}$ are
$h_{tt}^{(0)}=2,h_{rr}^{(0)}=2$, and we assume the $h_{\mu\nu}^{(1)}$
above are constants. We find that we can solve the linearised
equations of motion to first order in $\mu^2$ for this ansatz by
setting $h^{(1)}_{tt}=4,h^{(1)}_{tz}=3,h^{(1)}_{zz}=2,
h^{(1)}_{\theta\theta}=-1, h^{(1)}_{\phi\phi}=-1$. 

We can summarise these results in a more invariant fashion by saying
that for a plane wave background of the form
\begin{equation}
ds_{wave}^{2}=-dt^{2}+dz^{2}+dr^2+r^2(d\theta^2+\sin^2 \theta d\phi^2)
-\mu^{2}f(\theta,\phi)(dt+dz)^{2},
\end{equation}
a solution of the linearized equations of motion for a black string
source, to linear order in $\mu^2$, is 
\begin{eqnarray}
h_{tt} & = & \frac{2M}{r}+ 4M\mu^{2}r f(\theta, \phi),\\
h_{tz} & =& 3M\mu^{2}r f(\theta, \phi), \nonumber\\
h_{zz} & = & 2M\mu^{2}r f(\theta, \phi), \nonumber\\
h_{rr} & =& \frac{2M}{r},\nonumber \\ 
h_{\theta\theta} & =& - M\mu^{2}r^3 f(\theta, \phi), \nonumber\\
h_{\phi\phi} &=& - M \mu^2 r^3 \sin^2 \theta f(\theta,\phi).\nonumber
\end{eqnarray}
We would expect that this generalises straightforwardly to higher
dimensions. As in the black hole case, this demonstrates that these
solutions are not asymptotically plane wave, as the perturbation is
large compared to the background metric far from the source. 

\section{Near region analysis}
\label{near}

Having explored the behaviour in the intermediate region, where we can
use a linearised approximation about the plane wave background, we now
turn to the analysis in the region $r \ll \mu^{-1}$ near the black
hole or black string. In this region we can treat the plane wave as a
small perturbation of the black object, and the problem reduces to
linearised perturbations on the black hole or black string background,
with boundary conditions at large distances determined from the
previous solution in the intermediate region and a boundary condition
at the horizon determined by requiring regularity of the perturbed
solution there. We will find that there is no regular solution in the
black hole case. For the black string, we find a regular solution
which matches on to the solution we discussed above in the
intermediate region. We will focus on the analysis for the black hole
in four dimensions and the black string in five dimensions, as in the
previous section, but the same techniques can easily be applied in
higher dimensions. We will comment briefly on the extension of the
analysis to higher dimensions for the black hole case. 

\subsection{Black hole}
\label{nearbh}

We first study the near region of the black hole, treating the plane
wave as a perturbation. We will do the analysis in the lowest possible
dimension, $D=4$, even though there is a simple symmetry argument that
no regular solution exists in this case. The calculation is simplest
in this dimension, and it serves to illustrate the method of
calculation, which will be very similar in higher dimensions. 

Take the Schwarzschild black hole solution in four dimensions,
\begin{equation}
ds^{2}=-f(r)dt^{2}+\frac{dr^{2}}{f(r)}+r^{2}(d\theta^{2}+\sin^{2}\theta
d\phi^{2}),\label{eq:BH}
\end{equation}
with $f(r)=1-2M/r$. We want to find a solution of the source-free
linearised vacuum equations on this background which asymptotically
approaches the four-dimensional plane wave \eqref{4dpw}. This implies that
we want a perturbation $h_{\mu\nu}$ with asymptotic boundary conditions 
\begin{equation}
  \lim_{r \to \infty} h_{\mu\nu} dx^\mu dx^\nu 
  =-\frac{\mu^{2}r^{2}}{2}\sin^{2}\theta({\normalcolor
    e^{2i\phi}}-{\normalcolor e^{-2i\phi}})(dt+\cos\theta
  dr-r\sin\theta d\theta)^{2} + \ldots,\label{eq:h_BH_infinity}\end{equation}
where the $\ldots$ denotes terms going like $\mu^2 M^n$ for $n \neq
0$. These terms are suppressed relative to the leading term because
dimensional analysis tells us the mass will always appear in the
combination $M/r$. At linear order in $M$ the subleading terms at
large $r$ should match onto the results of the analysis in the
intermediate region obtained in the previous section. 

At the horizon, the boundary condition is that the solution be regular
there. Since the background metric is not regular at the horizon in
the Schwarzschild coordinate system we are using, this condition is
most easily applied by writing the perturbation in an orthonormal
frame. A suitable frame is $e^{(0)} = \sqrt{f(r)} dt$, $e^{(1)} =
f(r)^{-1/2} dr$, $e^{(2)} = r d\theta$, $e^{(3)} = r \sin \theta
d\phi$. Requiring that the components of the perturbation in the
orthonormal frame are regular at the horizon implies that we must
require that as $r \to 2M$, 
\begin{equation}
h_{tt} \sim (r-2M), \quad h_{t\mu} \sim (r-2M)^{1/2} \ \mathrm{ for }\
\mu
\neq t, \label{hbc}
\end{equation}
\begin{equation}
h_{rr} \sim (r-2M)^{-1}, \quad h_{r\mu} \sim (r-2M)^{-1/2}
\ \mathrm{ for }\ \mu \neq r, 
\end{equation}
\begin{equation}
h_{ij} \sim (r-2M)^0.
\end{equation}
These conditions can also be derived by requiring finiteness of
$h_{\mu\nu}$ in a coordinate system which is well-behaved at $r=2M$,
such as Kruskal coordinates.

Matching the leading term written in \eqref{eq:h_BH_infinity} and
imposing regularity at the horizon should determine the solution of
the perturbation equations uniquely. In fact, as we mentioned above,
we will find that there is no solution of the linearised perturbation
equations that satisfies these two boundary conditions.

For the black hole case, the analysis of the components on the sphere
is sufficiently complicated that it is useful to exploit the results
of \cite{Kodama:2003jz} on the spherical harmonic decomposition for perturbations
of Schwarzschild and rewrite the linearised equations of motion in
terms of gauge-invariant variables with respect to coordinate
transformations on the sphere. We therefore want to convert
\eqref{eq:h_BH_infinity} into boundary conditions for their
gauge-invariant perturbations. Let $a,b=t,r$ and $i,j=\theta,\phi.$
Then we have boundary conditions which are scalars $h_{ab}$, vectors
$h_{ai}$, and a tensor $h_{ij}$, for which the boundary condition only
has an $h_{\theta\theta}$ component. Following \cite{Kodama:2003jz} we expand the
perturbation in terms of harmonics on $S^{2}$: the scalar harmonics
\begin{equation}
\square S=-l(l+1)S,\quad l=0,1,2, \ldots , 
\end{equation}
the vector harmonics 
\begin{equation}
\square V_{i}=(-l(l+1)+1)V_{i},\quad l=1,2,3, \ldots ,
\end{equation}
with $D_{i}V^{i}=0$, and the transverse traceless tensor harmonics 
\begin{equation}
\square T_{ij}=(-l(l+1)+2)T_{ij},\quad l=2,3,4, \ldots, 
\end{equation}
with $D_{i}T_{j}^{i}=0,T_{i}^{i}=0$.  There are, however, no pure
tensor harmonics $T_{ij}$ on $S^{2}.$ We use the notation
$\square=D_{i}D^{i}$ for the d'Alembertian operator on $S^{2}$, where
$D_{i}$ is the covariant derivative with respect to the metric
$\gamma_{ij}$ on the unit two-sphere.

In terms of these harmonics, the scalar components of the perturbation
are
\begin{equation}
{\normalcolor h_{ab}=\sum_{l,m}{\displaystyle
    f_{ab}S_{l}^{m}}}.\label{eq:scalar}
\end{equation}
Note that here and hereafter we will omit the $l,m$ indices on the
coefficients $f_{ab}$ or equivalent in the general relations like this
for brevity. The vector perturbations are decomposed into their scalar
derived and pure vector components $h_{ai}=h_{ai}^{S}+h_{ai}^{V}$,
where
\begin{equation}
h_{ai}^{S}=r\sum_{l,m}f_{a}(-\frac{1}{k^{2}}D_{i}S_{l}^{m}),
\end{equation}
where $k^{2}=l(l+1)$, and 
\begin{equation}
h_{ai}^{V}=r\sum_{l,m}f_{a}^{V}(V_{l}^{m})_{i}.
\end{equation}
Similarly, the tensor part of the perturbation is decomposed into
scalar derived, vector derived and pure tensor components
$h_{ij}=h_{ij}^{S}+h_{ij}^{V}+h_{ij}^{T}$, where
\begin{equation}
h_{ij}^{S}=2r^{2}\sum_{l,m}(H_{L}\gamma_{ij}S_{l}^{m}+H_{T}S_{ij}),
\end{equation}
where
$S_{ij}=\frac{1}{k^{2}}D_{i}D_{j}S_{l}^{m}+\frac{\gamma_{ij}}{2}S_{l}^{m}$,
\begin{equation}
h_{ij}^{V}=2r^{2}\sum_{l,m}H_{L}^{V}V_{ij},
\end{equation}
where $V_{ij}=-\frac{1}{2k_{V}^{2}}(D_{i}V_{j}+D_{j}V_{i})$ with
$k_{V}^{2}=l(l+1)-1,$ and 
\begin{equation}
h_{ij}^{T}=2r^{2}\sum_{l,m}H_{L}^{T}T_{ij}.
\end{equation}
There are, however, no pure tensor harmonics $T_{ij}$ on $S^{2}.$

Thus, to determine the boundary conditions for the gauge invariant
variables, we must apply this expansion to \eqref{eq:h_BH_infinity}
and find the asymptotic values for the unknown expansion
coefficients. For scalar perturbations this is
straightforward. Substituting \eqref{eq:h_BH_infinity} into
\eqref{eq:scalar} we are able to read off that
\begin{equation}
\lim_{r \to \infty} (f_{tt})_{2}^{\pm2} =-\frac{\mu^{2}r^{2}}{2},\quad
\lim_{r \to \infty} (f_{rr})_{2}^{\pm2}=-\frac{\mu^{2}r^{2}}{14},
\end{equation}
\begin{equation}
\lim_{r \to \infty} (f_{tr})_{3}^{\pm2} =-\mu^{2}r^{2},\quad
\lim_{r \to \infty} (f_{rr})_{4}^{\pm2} =-\frac{\mu^{2}r^{2}}{14}.
\end{equation}
We now turn our attention to the vector perturbations. Since $D^{i}V_{i}=0$
we have $D^{i}h_{ai}^{V}=0,$ so
\begin{equation}
D^{i}h_{ai}=D^{i}h_{ai}^{S}=r\sum_{l,m}f_{a}S_{l}^{m},
\end{equation}
where we have used $D^{i}D_{i}S=-k^{2}S$. Explicit computation gives
us the boundary conditions for the scalar derived vector coefficients,
\begin{equation}
\lim_{r \to \infty} (f_{r})_{2}^{\pm2}=-\frac{\mu^{2}r^{2}}{7},\quad
\lim_{r \to \infty} (f_{t})_{3}^{\pm2}=2\mu^{2}r^{2},\quad
\lim_{r \to \infty} (f_{r})_{4}^{\pm2}=\frac{5\mu^{2}r^{2}}{14}.
\end{equation}
To find the pure vector coefficients we write
\begin{equation}
h_{ai}^{V}=h_{ai}-h_{ai}^{S}=h_{ai}+r\sum_{l,m}f_{a}\frac{1}{k^{2}}D_{i}S_{l}^{m}=r\sum_{l,m}f_{a}^{V}(V_{l}^{m})_{i}.
\end{equation}
Again, by explicit computation we find,
\begin{equation}
\lim_{r \to \infty} (f_{t}^{V})_{2}^{\pm2}=\frac{\mu^{2}r^{2}}{3},\quad
\lim_{r \to \infty} (f_{r}^{V})_{3}^{\pm2}=\frac{\mu^{2}r^{2}}{6}.
\end{equation}
Finally we consider the tensor perturbations. We can write 
\begin{equation}
h^{i}{}_{i}=(h^{S})^{i}{}_{i}=4r^{2}\sum_{l,m}H_{L}S_{l}^{m},
\end{equation}
where we have used $D^{i}V_{i}=0,T^{i}{}_{i}=0,S^{i}{}_{i}=0$ and
$\gamma^{i}{}_{i}=2.$ This allows us to easily show that
\begin{equation}
\lim_{r \to \infty} (H_{L})_{2}^{\pm2}=
-\frac{3\mu^{2}r^{2}}{28}, \quad \lim_{r \to \infty}
(H_{L})_{4}^{\pm2} =\frac{\mu^{2}r^{2}}{56}.
\end{equation}
To find the scalar derived transverse modes we will need the following
results,
\begin{equation}
D^{i}D^{j}V_{ij}=0,
\end{equation}
\begin{equation}
D^{i}D^{j}S_{ij}=\frac{(k^{2}-2)}{2}S,
\end{equation}
which are proved in an appendix. Using the above results
along with $D^{i}T_{ij}=0,$ we find
\begin{eqnarray}
D^{i}D^{j}h_{ij} & = & D^{i}D^{j}h_{ij}^{S}\\
 & = & 2r^{2}\sum_{l,m}(-k^{2}H_{L}S_{l}^{m}+H_{T}D^{i}D^{j}S_{ij})
 \nonumber \\
 & = &
 2r^{2}\sum_{l,m}(-k^{2}H_{L}+H_{T}\frac{(k^{2}-2)}{2})S. \nonumber 
\end{eqnarray}
We can now show that
\begin{equation}
\lim_{r \to \infty} (H_{T})_{2}^{\pm2} =-\frac{\mu^{2}r^{2}}{7},\quad
\lim_{r \to \infty} (H_{T})_{4}^{\pm2}=-\frac{5\mu^{2}r^{2}}{12\cdot7}.
\end{equation}
To find the vector derived transverse modes we will use the identities
\begin{equation}
D^{i}S_{ij}=-\frac{1}{2k^{2}}(k^{2}-2)D_{j}S,
\end{equation}
and
\begin{equation}
D^{i}V_{ij}=\frac{1}{2k_{V}^{2}}(k_{V}^{2}-1)V_{j},
\end{equation}
which we also prove in an appendix. Since $D^{i}T_{ij}=0$, we have
\begin{equation}
D^{i}h_{ij}=D^{i}h_{ij}^{S}+D^{i}h_{ij}^{V},
\end{equation}
and using the results above we can write this as
\begin{equation}
D^{i}h_{ij}=2r^{2}\sum_{l,m}(H_{L}-\frac{1}{2k^{2}}(k^{2}-2)H_{T})D_{j}S+2r^{2}\sum_{l,m}H_{T}^{V}\frac{1}{2k_{V}^{2}}(k_{V}^{2}-1)V_{j}.
\end{equation}
We are now able to show that
\begin{equation}
\lim_{r \to \infty} (H_{T}^{V})_{3}^{\pm2}
=-\frac{11\mu^{2}r^{2}}{12}.
\end{equation}
Using Maple we find that $h_{ij}=h_{ij}^{S}+h_{ij}^{V},$ so there
are no pure tensor perturbations as expected. 

We now want to translate this into boundary conditions for the
gauge-invariant variables introduced in \cite{Kodama:2003jz}. For vector
perturbations the gauge-invariant variable is
\begin{equation}
F_{a}=f_{a}^{V}+\frac{r}{k_{V}^{2}}D_{a}H_{T}^{V}.
\end{equation}
For $l=2$, $\lim_{r \to \infty}
(f_{t}^{V})_{2}^{\pm2}=\frac{\mu^{2}r^{2}}{3},$ so
$\lim_{r \to \infty} F_{t}=\frac{\mu^{2}r^{2}}{3},$ and we have
$F_{a}=r^{-1}\epsilon_{ab}D^{b}(r\Phi)$ \cite{Kodama:2003jz}, so the boundary
condition for the vector master function $\Phi_{l=2}^{V}$ is
\begin{equation}
\lim_{r \to \infty} \Phi_{l=2}^{V} =\frac{\mu^{2}r^{3}}{12}.
\end{equation}
For $l=3$, $\lim_{r \to \infty} (f_{r}^{V})_{3}^{\pm2} =\frac{\mu^{2}r^{2}}{6}$
and $\lim_{r \to \infty} (H_{T}^{V})_{3}^{\pm2} =-\frac{11\mu^{2}r^{2}}{12}$
so $F_{a}=0$: this mode is pure gauge. This is as we might expect: the
$r^2$ behaviour of the plane wave is typical of an $l=2$ spherical
harmonic, so the higher $l$ modes that seem to appear in our
decomposition of the mode in terms of spherical harmonics ought to be
pure gauge. 

For scalar perturbations, the gauge-invariant variables are
\cite{Kodama:2003jz}
\begin{eqnarray}
F & = & H_{L}+\frac{1}{2}H_{T}+\frac{1}{r}(D^{a}r)X_{a}\\
F_{ab} & = & f_{ab}+D_{a}X_{b}+D_{b}X_{a}
\end{eqnarray}
with
\begin{equation}
X_{a}=\frac{r}{k^{2}}(f_{a}+rD_{a}H_{T}).
\end{equation}
The master variable $\Phi$ is
\begin{equation}
\Phi=\frac{2\tilde{Z}-r(X+Y)}{4},
\end{equation}
with
\begin{eqnarray}
X & = & F_{t}^{t}-2F\\
Y & = & F_{r}^{r}-2F\\
\tilde{Z} & = & 0.
\end{eqnarray}
For $l=2$ perturbations direct substitution gives us $\lim_{r \to
  \infty} X =\mu^{2}r^{2},Y=0,\tilde{Z}=0$, hence the
boundary condition on $\Phi$ is
\begin{equation}
\lim_{r \to \infty} \Phi_{l=2}^{S} = -\frac{\mu^{2}r^{3}}{4}.
\end{equation}
For the $l=3$ and $l=4$ modes we find the gauge-invariant variables
$F$ and $F_{ab}$ are zero, so these modes are pure gauge as expected.
Thus, we are left with two non-trivial modes, the $l=2$ scalar and
the $l=2$ vector modes. 

Having established which modes are non-zero
and their boundary conditions we consider the bulk solution. For the
vector mode the equation for the master field is \cite{Kodama:2003jz}
\begin{equation}
\partial_{r}((1-\frac{2M}{r})\partial_{r}\Phi)-\frac{1}{r^{2}}[l(l+1)-3\cdot\frac{2M}{r}]\Phi=0.\label{eq:Master}
\end{equation}
The boundary condition is $\lim_{r \to \infty} \Phi_{l=2}^{V}
=\frac{\mu^{2}r^{3}}{12}$, therefore we set $\Phi=r^{3}\psi.$ This
allows us to reduce the master equation \eqref{eq:Master} to
\begin{equation}
\partial_{r}(r^{6}(1-\frac{2M}{r})\partial_{r}\psi)=0.
\end{equation}
which has solution
\begin{equation}
\psi=a\left(\frac{1}{8Mr^{4}}+\frac{1}{12M^{2}r^{3}}+\frac{1}{16M^{3}r^{2}}+\frac{1}{16M^{4}r}+\frac{1}{32M^{5}}\ln(1-\frac{2M}{r})\right)+b.
\end{equation}
Solutions with $a\neq0$ are clearly not regular at $r=2M,$ therefore
the solution for the vector master field is $\Phi^{V}=br^{3}$. The
boundary condition at large $r$ then requires
$b=\frac{\mu^2}{12}$. However, the boundary condition at the horizon
\eqref{hbc} requires that $h_{tt}$ and $h_{ti}$ vanish at the
horizon. This implies that $f_{t}^{V}$ and hence $F_{t}^{V}$ also
vanish at the horizon.  Finally $F^{t}=r^{-1}D_{r}(r\Phi)$ implies
that $\Phi$ too must vanish at the horizon, which would require
$b=0$. Hence, there is no solution which satisfies the boundary
conditions at both the horizon and infinity.

Thus, there is no regular solution describing a four-dimensional black
hole in the plane wave background \eqref{4dpw}. In fact, this is not a
surprising result in four dimensions; the rigidity theorem
\cite{Hawking:1971vc} shows that regular black holes must be static or
stationary axisymmetric, and the plane wave \eqref{4dpw} is not static
and does not preserve a $U(1)$ symmetry. Thus, the plane wave
perturbation breaks too many of the symmetries of the black hole for a
regular deformed black hole solution to be possible.

One might hope to avoid this problem by considering a non-vacuum plane
wave solution. We can for example consider in four dimensions the
electromagnetic plane wave
\begin{equation}
ds^2_{wave} = -dt^2 + dx^2 + dy^2+ dz^2 - \mu^2 (x^2+y^2) (dt+dz)^2
\end{equation}
supported by the electric flux 
\begin{equation}
F = 2 \mu (dt+dz) \wedge dx. 
\end{equation}
This is also interesting as a simplified model of the maximally
supersymmetric plane wave of \cite{Blau:2002dy}. Here, the metric
perturbation preserves a $U(1)$ symmetry, but this is broken by the
gauge field, and as a result, we again do not expect to find a regular
black hole solution. In this case, the problem is that the equation of
motion for the gauge field on the Schwarzschild black hole background
has no solution which is regular on the horizon and satisfies the
boundary condition at large $r$.

If we consider the situation in higher dimensions, the above rigidity
argument does not apply, but there is still no regular solution. Take
for example a six-dimensional Schwarzschild black hole and add as a
perturbation the six-dimensional vacuum plane wave
\begin{equation}
ds^2_{wave} = -dt^2 + dv^2+dw^2+dx^2+dy^2+dz^2- \mu^2(v^2+w^2-x^2-y^2)
(dt+dz)^2. 
\end{equation}
This clearly preserves two $U(1)$ isometries, in the $x-y$ and $v-w$
planes. However, if we rewrite this in spherical polars, there is
again an $l=2$ vector part to the perturbation in the decomposition
into spherical harmonics. The analysis is very similar to the above
four-dimensional case, and it is not possible to find a solution for
the vector part of the perturbation that satisfies the plane wave
boundary conditions at large distances and the regularity condition on
the event horizon. In this case, the plane wave preserves two $U(1)$
isometries on the $S^4$ surrounding the black hole, so the above
argument does not apply; a regular deformed black hole solution would
not violate the conditions of \cite{Hollands:2006rj}. This problem
seems to be very general. In all cases we have explored in the vacuum
Einstein equations, the plane wave has a vector part in the spherical
harmonic decomposition, and it is not possible to find a regular
perturbation of the black hole which satisfies the plane wave boundary
condition.  It would be interesting to understand the physical origins
of this restriction further.

\subsection{Black string}

We next study the near horizon region of the black string, treating the
plane wave as a perturbation. The background is the five-dimensional
black string solution
\begin{equation}
ds^{2}=-f(r)dt^{2}+\frac{dr^{2}}{f(r)}+r^{2}(d\theta^{2}+\sin^{2}\theta
d\phi^{2})+dz^{2},\label{eq:BS}
\end{equation}
 with $f(r)=1-2M/r$. We want to find a solution of the source-free
linearised vacuum equations on this background which asymptotically
approaches the five-dimensional plane wave \eqref{5dwave}. This implies that
we want a perturbation $h_{\mu\nu}$ with asymptotic boundary
conditions 
\begin{equation}
\lim_{r \to \infty}  h_{\mu\nu} dx^\mu dx^\nu =-\mu^{2}r^{2}[ \alpha
(1-3\cos^{2}\theta) + \beta \sin^{2}\theta(\cos^{2}\phi-\sin^{2}\phi)]
(dt+dz)^{2} + \ldots, 
\end{equation}
where the $\ldots$ denotes terms going like $\mu^2 M^n$ for $n \neq
0$. These terms are suppressed relative to the leading term because
dimensional analysis tells us the mass will always appear in the
combination $M/r$. 

As in the analysis in the intermediate region, we will deal with the
$\alpha$ and $\beta$ components separately. It will turn out that the
analysis is identical in these two cases. In terms of the spherical
harmonic analysis on the two-sphere, these are scalar-type
perturbations, which excite the $l=2,m=0$ and $l=2,m=2$ harmonic modes
respectively. In the linearised theory, we can assume that the
perturbation has only these modes turned on. Since only scalar-type
modes are excited, the analysis on the sphere is fairly simple, and we
will follow he similar analysis by Emparan et al
\cite{Emparan:2007wm}.

The boundary conditions, and hence the perturbation, are invariant
under simultaneously taking $t\to-t$, $z\to-z$, so the only modes we
need to consider are $h_{tt}$, $h_{tz}$, $h_{zz}$, $h_{rr}$, and the
longitudinal and transverse scalar-derived perturbations on the
sphere.

We first consider only the $l=2,m=0$ perturbation
(we set $\beta=0$). Assuming that only this spherical
harmonic is excited, we can write the perturbation as 
\begin{equation}
h_{tt}=\alpha(1-3\cos^{2}\theta)a(r),\quad
h_{tz}=\alpha(1-3\cos^{2}\theta)b(r),\quad h_{zz}=\alpha(1-3\cos^{2}\theta)c(r),
\end{equation}
\begin{equation}
h_{rr}=\alpha\frac{(1-3\cos^{2}\theta)}{(1-2M/r)}f(r),
\end{equation}
\begin{equation}
h_{\theta\theta}=\alpha r^{2}[(1-3\cos^{2}\theta)g(r)-3\sin^{2}\theta h(r)],
\end{equation}
\begin{equation}
h_{\phi\phi}=\alpha r^{2}\sin^{2}\theta[(1-3\cos^{2}\theta)g(r)+3\sin^{2}\theta
h(r)].
\end{equation}
Note that $g(r)$ is the coefficient of the longitudinal mode on the
sphere, and $h(r)$ is the coefficient of the transverse mode on the
sphere.
As in \cite{Emparan:2007wm}, there is a remaining coordinate freedom,
under 
\begin{equation} r
\to
  r+\gamma(r)(1-3\cos^{2}\theta),\quad\theta\to\theta+6\beta(r)\cos\theta\sin\theta,
\end{equation}
with 
\begin{equation}
  \beta'(r)=-\frac{\gamma(r)}{r(r-2M)},\quad\gamma(2M)=0.
\end{equation}

Similarly, for the $l=2,m=2$ pertubation (obtained by setting $\alpha=0$),
we define
\begin{equation}
h_{tt}=\beta\sin^{2}\theta(\cos^{2}\phi-\sin^{2}\phi)a(r),\quad
h_{tz}=\beta\sin^{2}\theta(\cos^{2}\phi-\sin^{2}\phi)b(r),
\end{equation}
\begin{equation}
h_{zz}=\beta\sin^{2}\theta(\cos^{2}\phi-\sin^{2}\phi)c(r),
\end{equation}
\begin{equation}
h_{rr}=\beta\frac{\sin^{2}\theta(\cos^{2}\phi-\sin^{2}\phi)}{(1-2M/r)}f(r),
\end{equation}
\begin{equation}
h_{\theta\phi}=\beta r^{2}\sin\theta\cos\theta\sin\phi\cos\phi h(r),
\end{equation}
\begin{equation}
h_{\theta\theta}=\beta r^{2}[\sin^{2}\theta(\cos^{2}\phi-\sin^{2}\phi)g(r)-(\cos^{2}\theta+1)(\cos^{2}\phi-\sin^{2}\phi)h(r)],
\end{equation}
\begin{equation}
h_{\phi\phi}=\beta r^{2}\sin^{2}\theta[\sin^{2}\theta(\cos^{2}\phi-\sin^{2}\phi)g(r)+(\cos^{2}\theta+1)(\cos^{2}\phi-\sin^{2}\phi)h(r)].
\end{equation}
Now we have remaining coordinate freedom under 
\begin{equation}
r\to r+\gamma(r)\sin^{2}\theta(\cos^{2}\phi-\sin^{2}\phi),
\end{equation}
\begin{equation}
\theta\to\theta+2\beta(r)\sin\theta\cos\theta(\cos^{2}\phi-\sin^{2}\phi),
\end{equation}
\begin{equation}
\phi\to\phi-4\beta(r)\sin^{2}\theta\cos\phi\sin\phi,
\end{equation}
with 
\begin{equation} \label{gtbc}
\beta'(r)=-\frac{\gamma(r)}{r(r-2M)},\quad\gamma(2M)=0.
\end{equation}

We find both coordinate transformations produce identical shifts 
\begin{equation}
a(r)\to a(r)-\frac{2M}{r^{2}}\gamma(r),\quad f(r)\to
f(r)+\left(2\gamma'-\frac{2M}{r}\frac{\gamma(r)}{r-2M}\right),
\end{equation}
\begin{equation}
g(r)\to g(r)+\frac{2}{r}\gamma(r)-6\beta(r),\quad h(r)\to
h(r)+2\beta(r),
\end{equation}
while $b(r)$ and $c(r)$ are unchanged. 

We want to consider combinations which are invariant under these
coordinate transformations. $B=b(r)$ and $C=c(r)$ are already
invariant. We define in addition  
\begin{equation}
A=a(r)+\frac{M}{r}(g(r)+3h(r)),\label{eq:A}
\end{equation}
\begin{equation}
F=f(r)-\frac{d}{dr}\left(r(g(r)+3h(r))\right)+\frac{M(g(r)+3h(r))}{(r-2M)},\label{eq:F}
\end{equation}
\begin{equation}
H'=\frac{dh}{dr}+\frac{g(r)+3h(r)}{(r-2M)}.\label{eq:H}
\end{equation}
Note that in this section, primes denote derivatives with respect to
$r$. As in \cite{Emparan:2007wm}, the constant part of $h(r)$ can be
fixed using the constant part of $\beta(r)$. Using the gauge-invariant
combinations basically amounts to setting $g(r)=-3h(r)$, which can be
achieved for $r \neq 2M$ by an appropriate choice of gauge. Because of
the boundary condition in \eqref{gtbc}, $g(2M)+3h(2M)$ is
gauge-invariant. It will however not be determined by solving the
equations of motion for the above gauge-invariant variables, and will
have to be separately specified. It will turn out to be determined by
requiring regularity of the solution at the horizon.

For either $\alpha=0$ or $\beta=0$, substituting into the linearised
Einstein equations gives the same system of equations for the unknown
functions $A,B,C,F,H'$ (keeping terms up to $\mathcal{O}(\mu^{2})$),
\begin{eqnarray}
R_{tt}^{(1)} & \propto & r^{2}(r-2M)^{2}A''+r(r-2M)(2r-5M)A'-M(r-2M)^{2}C'\label{tteq}\\
 &  & -(6r(r-2M)-2M^{2})A+M(r-2M)^{2}F'+6M(r-2M)^{2}H',\nonumber 
\end{eqnarray}
\begin{equation}
R_{tz}^{(1)}\propto r(r-2M)B''+2(r-2M)B'-6B,\label{beq}
\end{equation}
\begin{equation}
R_{zz}^{(1)}\propto r(r-2M)C''+2(r-M)C'-6C,\label{ceq}
\end{equation}
\begin{eqnarray}
R_{rr}^{(1)} & \propto & r^{2}(r-2M)^{2}A''-rM(r-2M)A'+2M(2r-3M)A\label{rreq}\\
 &  & -r(r-2M)^{3}C''-M(r-2M)^{2}C'+(2r-3M)(r-2M)^{2}F'\nonumber \\
 &  & +6(r-2M)^{2}F+6r(r-2M)^{3}H''+6(2r-3M)(r-2M)^{2}H',\nonumber 
\end{eqnarray}
\begin{eqnarray}
R_{r\theta}^{(1)} & \propto & -r^{2}(r-2M)A'+r(r-M)A+r(r-2M)^{2}C'-(r-2M)^{2}C\label{rtheq}\\
 &  & -(r-2M)(r-M)F-r(r-2M)^{2}H',\nonumber 
\end{eqnarray}
\begin{eqnarray}
R_{\theta\theta}^{(1)}+\frac{1}{\sin^{2}\theta}R_{\phi\phi}^{(1)} & \propto & r(r-2M)A'-(3r+2M)A-(r-2M)^{2}C'+3(r-2M)C\nonumber \\
 &  & +(r-2M)^{2}F'+5(r-2M)F+3r(r-2M)^{2}H''\\
 &  & +6(2r-3M)(r-2M)H',\nonumber 
\end{eqnarray}
\begin{equation}
R_{\theta\theta}^{(1)}-\frac{1}{\sin^{2}\theta}R_{\phi\phi}^{(1)}\propto-rA+(r-2M)C+(r-2M)F+r(r-2M)^{2}H''+2(r-M)(r-2M)H'.\label{heq}
\end{equation}
In fact it is easy to show that the linearised Einstein
equations must be the same for both modes. The perturbation involves
some $l=2$ scalar harmonic, let's call it $S$ , so
\begin{equation}
h_{ab}=f_{ab}(r)S,\quad h_{ai}=f_{a}(r)\nabla_{i}S,\quad
h_{ij}=f(r)Sg_{ij}+f^{\prime}(r)\nabla_{i}\nabla_{j}S,
\end{equation}
where $i,j$ are coordinates on the two-sphere and $a,b=t,r,z$. Then
the first order Ricci tensor constructed from the second covariant
derivatives of $h_{\mu\nu}$ will also depend on angular coordinates
only through $S$ and its derivatives. Using
$\nabla_{i}\nabla^{i}S=-6S$ and the fact that the sphere is an
Einstein space, so $R_{ij}=g_{ij}$, one can eliminate extra
derivatives of $S$, to leave us with
\begin{equation}
  R_{ab}^{(1)}=\epsilon_{ab}(r)S,\quad
  R_{ai}^{(1)}=\epsilon_{a}(r)\nabla_{i}S,\quad
  R_{ij}^{(1)}=\epsilon(r)Sg_{ij}+\epsilon^{\prime}(r)\nabla_{i}\nabla_{j}S.
\end{equation}
Hence, the resulting equations
$\epsilon_{ab}(r)=\epsilon_{a}(r)=\epsilon(r)=\epsilon^{\prime}(r)=0$
are independant of whether $S$ is in the $m=0$ or $m=2$ mode. Thus,
solving the equations (\ref{tteq}-\ref{heq}) will give us the general
solution for the perturbation in the near-horizon region for both
modes. 

The boundary conditions at large $r$ imply that at order $M^0$, $a(r),
b(r), c(r) \to -\mu^2 r^2$, and $f(r)$, $g(r), h(r)$ have no $\mu^2
M^0$ term. This implies that 
\begin{equation}
A, B, C \to - \mu^2 r^2,  
\end{equation}
and $F$ and $H'$ have no $\mu^2 M^0$ term. Regularity at the horizon
requires $a(r) \propto (r-2M)$, $b(r) \propto \sqrt{r-2M}$, and the
other functions $c(r), f(r), g(r)$ and $h(r)$ are required to be
finite there. In terms of the gauge-invariant combinations, these
boundary conditions are best expressed in terms of the alternative
combinations 
\begin{equation}
\bar{A} = A - \frac{M}{r}(r-2M) H', \quad \bar{F} = F - M H'. 
\end{equation}
The conditions for regularity at the horizon are then that $\bar{A}
\to 0$, $\bar{F}$ is finite, and $H'$ is allowed to diverge like
$(r-2M)^{-1}$.

We now want to solve this system of equations. We see that there are
two decoupled equations, \eqref{beq} and \eqref{ceq}. The solutions
of these satisfying our boundary conditions are 
\begin{equation}
B(r)=-\mu^{2}(r-M)(r-2M)\label{eq:B}
\end{equation}
and 
\begin{equation}
C(r)=-\mu^{2}(r^{2}-2Mr+\frac{2}{3}M^{2}).\label{eq:C}
\end{equation}
It is also convenient to subtract a multiple of \eqref{ceq} from
\eqref{rreq} to simplify it to 
\begin{eqnarray}
0 & = & r^{2}(r-2M)^{2}A''-rM(r-2M)A'+2M(2r-3M)A\\
 &  & +(2r-5M)(r-2M)^{2}C'-6(r-2M)^{2}C+(2r-3M)(r-2M)^{2}F'\nonumber \\
 &  & +6(r-2M)^{2}F+6r(r-2M)^{3}H''+6(2r-3M)(r-2M)^{2}H'.\nonumber 
\end{eqnarray}

We first solve \eqref{heq} for $A$, 
\begin{equation}
A=\frac{(r-2M)}{r}\left[C+F+r(r-2M)H''+2(r-2M)H'\right],
\end{equation}
and then solve $R_{tt}^{(1)}-(r-2M)^{2}R_{zz}^{(1)}-R_{rr}^{(1)}$
for $F$, 
\begin{equation}
F=\frac{1}{6}\left[r(r-2M)^{2}H'''-2(r-2M)(r+2M)H''-2(5r-7M)H'-MC'\right].
\end{equation}
The remaining equations then need to be solved for $H'$. By combining
equations, we can obtain a second-order inhomogeneous equation for
$H'$, 
\begin{eqnarray}
-2r(r+M)(r-2M)^{2}H'''-2(4r^{2}+3rM-4M^{2})(r-2M)H''\label{eq:diff_eq_H}\\
+2(4r^{2}-13rM+4M^{2})H'=M[(r-2M)C'+6C].\nonumber 
\end{eqnarray}
It's useful to note at this point that if $M=0$, we have a solution
with $F=H'=0$ and $A=C=-\mu^{2}r^{2}$, which is precisely our original
plane wave.

The general solution of \eqref{eq:diff_eq_H} is 
\begin{equation}
H'=\frac{\mu^{2}}{3}(r-M)+c_{1}\frac{r^{2}-2M^{2}}{r-2M}+c_{2}\frac{\left[-6rM(r+M)+4M^{3}+(6rM^{2}-3R^{3})\ln(1-2M/r)\right]}{r(r-2M)}.
\end{equation}
This then satisfies all of the equations. To get a solution which
is both regular and has the correct asymptotics, i.e. has $A\to-\mu^{2}r^{2}$
at large $r$, we need to take $c_{1}=-\frac{1}{3}\mu^{2}$ and $c_{2}=0.$
We find 
\begin{equation}
H'=-\frac{\mu^{2}M}{3}\frac{3r-4M}{r-2M},
\end{equation}
and
\begin{equation}
A=-\mu^{2}\left[r^{2}-4rM+\frac{16}{3}M^{2}-2\frac{M^{3}}{r}\right],\quad
F=\frac{2\mu^{2}M}{3}\frac{3r^{2}-9rM+5M^{2}}{r-2M}.\label{eq:HAF}
\end{equation}
In terms of the alternative combinations $\bar{A}$, $\bar{F}$, 
\begin{equation}
\bar{A} = -\mu^2(r-2M)\left[r-2M + \frac{M^2}{3r} \right], \quad
\bar{F} = \mu^2 M (2r-M). 
\end{equation}
Thus, this solution satisfies the regularity conditions at the
horizon. Regularity of the original functions $a(r), f(r),g(r),h(r)$
at $r=2M$ further requires us to choose 
\begin{equation}
g(2M)+3h(2M)=-\frac{2\mu^{2}M^{2}}{3}.
\end{equation}

We now match the near horizon and intermediate region solutions in the
intermediate region $\mu^{-1}\gg r\gg M,$ where both approximations
are valid. The contribution from the black string background is
\begin{equation}
ds_{NR,BG}^{2}\approx-(1-\frac{2M}{r})dt^{2}+(1+\frac{2M}{r})dr^{2}+r^{2}(d\theta^{2}+\sin^{2}\theta
d\phi^{2})+dz^{2}.
\end{equation}
We must now find the unknown functions $a(r),b(r),c(r),f(r),g(r),h(r)$
in this region to obtain the contribution from the perturbation. In
addition to the solutions \eqref{eq:B}, \eqref{eq:C} and
\eqref{eq:HAF} we must make a choice of gauge. We choose
$g+3h=-M\mu^{2}r$ in order to make the $rr$-component of the
perturbation vanish, matching our gauge choice in the intermediate
region solution.  We find, keeping just the terms up $\mathcal{O}(M)$ and
$\mathcal{O}(\mu^{2})$,
\begin{equation}
a(r)\approx -\mu^{2}(r^{2}-4Mr),\quad b(r)\approx -\mu^{2}(r^{2}-3Mr),\quad
c(r)\approx -\mu^{2}(r^{2}-2Mr),
\end{equation}
\begin{equation}
f(r)\approx 0,\quad g(r)\approx -M\mu^{2}r,\quad h(r)\approx 0.
\end{equation}
Hence the near region perturbation is
\begin{eqnarray}
ds_{NR,P}^{2} & \approx & (\alpha(1-3\cos^{2}\theta)+\beta\sin^{2}\theta(\cos^{2}\phi-\sin^{2}\phi))\times\label{eq:-1}\\
 &  &
 (-\mu^{2}r^{2}(dt+dz)^{2}+M\mu^{2}r(4dt^{2}+6dtdz+2dz^{2}-r^{2}(d\theta^{2}+\sin^{2}\theta
 d\phi^{2}))).\nonumber 
\end{eqnarray}

In the intermediate region the plane wave background is,
\begin{eqnarray}
ds_{IR,BG}^{2}&=&-dt^{2}+dz^{2}+dr^{2}+r^{2}(d\theta^{2}+\sin^{2}\theta
d\phi^{2}) \\
&&-\mu^{2}r^{2}(\alpha(1-3\cos^{2}\theta)+\beta\sin^{2}\theta(\cos^{2}\phi-\sin^{2}\phi))(dt+dz)^{2}. \nonumber
\end{eqnarray}
From section \ref{intbs}, the perturbation due to the black string is
\begin{eqnarray}
ds_{IR,P}^{2} & = & \frac{2M}{r}dt^{2}+\frac{2M}{r}dr^{2}+M\mu^{2}r(\alpha(1-3\cos^{2}\theta)+\beta\sin^{2}\theta(\cos^{2}\phi-\sin^{2}\phi))\times\nonumber \\
 &  & (4dt^{2}+6dtdz+2dz^{2}-r^{2}(d\theta^{2}+\sin^{2}\theta
 d\phi^{2})).\label{eq:}
\end{eqnarray}
Thus the solution constructed in the near region
$ds_{NR}^{2}=ds_{NR,BG}^{2}+ds_{NR,P}^{2}$ agrees with the solution
constructed in the intermediate region
$ds_{IR}^{2}=ds_{IR,BG}^{2}+ds_{IR,P}^{2}$ to the relevant order. This
gives us an approximate solution describing a black string in a plane
wave, valid when the size of the black string is small compared to the
curvature scale of the wave, $r_+ \ll \mu^{-1}$.

As in \cite{Emparan:2007wm}, the perturbation does not affect the
thermodynamic properties of the black hole at this order. The area of
the horizon cannot be affected at this order because the perturbation
is entirely in an $l=2$ mode, which deforms the shape of the $S^2$ but
does not change its area. The temperature cannot be affected because
it is constant over the horizon. Since the perturbation is an $l=2$
mode, it will vanish at some point on the horizon, whence the
temperature at that point must be unaffected, and since it is
constant, it must be unchanged over the whole horizon.

\section{Conclusions}
\label{concl}

In this paper, we have attempted to construct solutions describing
black holes and black strings in plane wave backgrounds using the
matched asymptotic expansion method. We have found that it is not
possible to construct a regular black hole solution. In the
approximation where the wave is thought of as a linearised
perturbation on the black hole solution, we need a non-zero vector
part in the spherical harmonic decomposition on the sphere, and it is
not possible to make this vector part regular on the horizon. It would
be interesting to have a deeper physical understanding of this failure
of regularity. One might think that this is simply saying that the
plane wave is exerting a force on the black hole, so no stationary
solution exists. However, we do not believe this is the correct
interpretation of our result. The black hole was chosen to follow a
geodesic in the plane wave background, so there is no force on it at
leading order. Finite size effects can be analysed in the asymptotic
region using the classical effective field theory approach of
\cite{Goldberger:2004jt,Goldberger:2005cd,Kol:2007rx}. In this
approach, the work done by such finite size terms involves derivatives
of the long wavelength background fields along the black hole
world-line. Since our world-line is chosen to be an orbit of the
isometries of the background, the work done will vanish. Thus, we
would have expected the background to simply produce some deformation
of the horizon.

The regularity problem seems to be simply an inconsistency between the
symmetry structure of the black hole and the plane wave. In four
dimensions, the problem is that the solution will not be axisymmetric,
so there cannot be a regular black hole solution as all stationary
four-dimensional black holes are required to be axisymmetric
\cite{Hawking:1971vc}. In higher dimensions, however, stationary
axisymmetric solutions describing black holes in plane waves could in
principle exist, and the fact that our solutions are never regular is
somewhat mysterious. Further exploration of this issue is an interesting
project for the future. 

The importance of this problem is reinforced by the fact that the
failure of regularity here is a counter-example to the assumption in
\cite{Emparan:2009at} that satisfying the blackfold equations implies
horizon regularity. Understanding this issue in a more general context
is clearly important for the blackfolds program
\cite{Emparan:2009cs,Emparan:2009at}; in considering the embedding of
black branes in arbitrary backgrounds, we need to understand when the
resulting deformation of the near-horizon region will preserve the
regularity of the event horizon. Clearly we must require that the
embedding of the blackfold in the background spacetime preserves
enough symmetry to satisfy the rigidity theorems of
\cite{Hawking:1971vc,Hollands:2006rj}. Our higher-dimensional examples
indicate that this is a necessary but not a sufficient
condition. Identifying sufficient conditions is an important general
problem.

We successfully constructed an approximate solution describing a black
string in an asymptotically vacuum plane wave background in 5
dimensions. It would clearly be interesting to extend this work to
find black string solutions in backgrounds which asymptote to
maximally supersymmetric plane waves. It should be straightforward to
extend our calculation to this case.

Our analysis has also led to an interesting general result; the effect
of localised objects in a plane wave background is not small, even far
from the source. The usual $1/r^{d-1}$ falloff associated with a
localised object in $d+1$ spatial dimensions is offset by the $\mu^2
r^2$ factors coming from the plane wave background. As a result, we
find that the ``perturbation'' due to the source is larger than the
background metric at sufficiently large $r$. This leads us to believe
that these solutions should not be thought of as ``asymptotically
plane wave'' spacetimes.

A definition of ``asymptotically plane wave'' was proposed in our
earlier work \cite{LeWitt:2008zx}, which allows the construction of a
well-behaved action principle. This still seems a useful
definition. However, from the present results it seems that the phase
space associated with those boundary conditions will not include
solutions describing localised sources in a plane wave background, so
it may not admit many physically interesting solutions. Understanding
the space of asymptotically plane wave spacetimes is clearly important
for attempts to construct a direct holographic duality directly for
plane waves, so we would like to understand this issue better.

Similar problems have arisen in AdS$_2$ spacetimes
\cite{Maldacena:1998uz}, where there are no finite-energy
asymptotically AdS$_2$ geometries, and in the study of near-horizon
extremal Kerr solutions (NHEK)
\cite{Guica:2008mu,Amsel:2009ev,Dias:2009ex}, where the space of
metrics which are asymptotically NHEK consists only of the NHEK
solution and solutions obtained from it by diffeomorphisms. It is
interesting to note that the plane waves, like AdS$_2$, have a
one-dimensional boundary
\cite{Berenstein:2002sa,Marolf:2002ye}. Perhaps the problem is that
there is in some sense ``not enough space'' near infinity to have
interesting asymptotically plane wave solutions. It would be
interesting to carry out a general analysis for asymptotically plane
wave solutions along the lines of that in
\cite{Amsel:2009ev,Dias:2009ex}. We leave this as a project for the
future.

\section*{Acknowledgements}
We are grateful for useful conversations with Veronika Hubeny and
Mukund Rangamani. This work was supported in part by the STFC.

\appendix

\section{Appendix}

In this appendix we prove some harmonic identities needed for our
analysis of black holes in the near horizon region. Definitions are
given in section \ref{nearbh}. We want to show that:
\begin{itemize}
\item $D^{i}D^{j}V_{ij}=0,$
\end{itemize}
Proof:
\begin{eqnarray}
D^{i}D^{j}V_{ij} & \propto & D^{i}D^{j}D_{i}V_{j}+D^{i}D^{j}D_{j}V_{i}\\
 & = & [D^{i},D^{j}]D_{i}V_{j}+2D^{j}D^{i}D_{i}V_{j} \nonumber \\
 & = & -R^{k}{}_{i}{}^{ij}D_{k}V_{j}-R^{k}{}_{j}{}^{ij}D_{i}V_{k}-2k_{V}^{2}D^{j}V_{j}\nonumber\\
 & = & R^{kj}D_{k}V_{j}-R^{ki}D_{i}V_{k}\nonumber\\
 & = & 0.\nonumber
\end{eqnarray}
\begin{itemize}
\item $D^{i}D^{j}S_{ij}=\frac{(k^{2}-2)}{2}S,$
\end{itemize}
Proof:
\begin{eqnarray}
D^{i}D^{j}S_{ij} & = & \frac{1}{k^{2}}D^{i}D^{j}D_{i}D_{j}S+\frac{1}{2}D^{j}D_{j}S\\
 & = & \frac{1}{k^{2}}D^{i}[D^{j},D_{i}]D_{j}S+\frac{1}{k^{2}}D^{i}D_{i}D^{j}D_{j}S+\frac{1}{2}D^{j}D_{j}S\nonumber\\
 & = & -\frac{1}{k^{2}}D^{i}(R^{k}{}_{j}{}^{j}{}_{i}S)+\frac{k^{2}}{2}S\nonumber\\
 & = & \frac{1}{k^{2}}D^{i}(R^{k}{}_{i}D_{k}S)+\frac{k^{2}}{2}S\nonumber
\end{eqnarray}
for $S^{2},$ $R_{ij}=\gamma_{ij}$ so
\begin{eqnarray}
D^{i}D^{j}S_{ij} & = & \frac{1}{k^{2}}D^{i}D_{i}S+\frac{k^{2}}{2}S\\
 & = & \frac{(k^{2}-2)}{2}S.\nonumber
\end{eqnarray}
\begin{itemize}
\item $D^{i}S_{ij}=-\frac{1}{2k^{2}}(k^{2}-2)D_{j}S,$
\end{itemize}
Proof:
\begin{eqnarray}
D^{i}S_{ij} & = & \frac{1}{k^{2}}D^{i}D_{i}D_{j}S+\frac{1}{2}D_{j}S\\
 & = & \frac{1}{k^{2}}[D^{i},D_{j}]D_{i}S+\frac{1}{k^{2}}D_{j}D^{i}D_{i}S+\frac{1}{2}D_{j}S \nonumber \\
 & = & -\frac{1}{k^{2}}R^{l}{}_{i}{}^{i}{}_{j}D_{l}S-\frac{1}{2}D_{j}S \nonumber \\
 & = & -\frac{1}{2k^{2}}(k^{2}-2)D_{j}S \nonumber 
\end{eqnarray}
\begin{itemize}
\item $D^{i}V_{ij}=\frac{1}{2k_{V}^{2}}(k_{V}^{2}-1)V_{j},$
\end{itemize}
Proof:
\begin{eqnarray}
D^{i}V_{ij} & = & -\frac{1}{2k_{V}^{2}}(D^{i}D_{i}V_{j}+D^{i}D_{j}V_{i})\\
 & = & \frac{1}{2}V_{j}-\frac{1}{2k_{V}^{2}}[D^{i},D_{j}]V_{i}-\frac{1}{2k_{V}^{2}}D_{j}D^{i}V_{i}\nonumber\\
 & = & \frac{1}{2}V_{j}+\frac{1}{2k_{V}^{2}}R^{k}{}_{i}{}^{i}{}_{j}V_{k}\nonumber\\
 & = & \frac{1}{2k_{V}^{2}}(k_{V}^{2}-1)V_{j}.\nonumber
\end{eqnarray}

\bibliographystyle{utphys}
\bibliography{planewave_blackholes}

\end{document}